\documentclass[twoside,11pt]{article}
\usepackage[a4paper, left = 1in, right = 1in, top = 1in, bottom = 1in]{geometry}
\usepackage{cite}
\usepackage{authblk}
\usepackage{amsmath,amssymb,amsfonts}
\usepackage{graphicx}
\usepackage{textcomp}
\usepackage{color}
\usepackage{hyperref}
\hypersetup{
	colorlinks=true,      
	urlcolor=blue,
	citecolor=black,
linkcolor=black,
filecolor=black
}
\usepackage{algorithm,algorithmic}
\usepackage{subfigure}
\usepackage{calrsfs}
\DeclareMathAlphabet{\pazocal}{OMS}{zplm}{m}{n}
\usepackage{bm}
\usepackage{bbm}
\usepackage[normalem]{ulem}
\usepackage{soul}
\usepackage{xargs}                      % Use more than one optional parameter in a new commands
\newcommand\redsout{\bgroup\markoverwith{\textcolor{red}{\rule[0.5ex]{2pt}{0.4pt}}}\ULon}
\usepackage{tablefootnote}
\begin{document}
\title{Simultaneous 12-Lead Electrocardiogram Synthesis using a Single-Lead ECG Signal: Application to Handheld ECG Devices}
\author[1]{Kahkashan Afrin}
\author[2]{Parikshit Verma}
\author[3]{Sanjay S. Srivatsa}
\author[1]{Satish T. S. Bukkapatnam}
\affil[1]{Department of Industrial and Systems Engineering, Texas A\&M University, College Station, TX 77843-3131, USA}
\affil[2]{Dell Inc., Austin, TX 78682, USA}
\affil[3]{Director, Heart Artery and Vein Center of Fresno, CA}
\affil[ ]{\textit {\{\href{mailto:afrin@tamu.edu}{afrin@tamu.edu},\href{mailto:parikshit\_verma@dell.com}{parikshit\_verma@dell.com},\href{mailto:sanjaysrivatsa@hotmail.com}{sanjaysrivatsa@hotmail.com},\href{mailto:satish@tamu.edu}{satish@tamu.edu}\}}}

\date{}
\maketitle

\begin{abstract}
Recent introduction of wearable single-lead ECG devices of diverse configurations has caught the intrigue of the medical community. While these devices provide a highly affordable support tool for the caregivers for continuous monitoring and to detect acute conditions, such as arrhythmia, their utility for cardiac diagnostics remains limited. This is because clinical diagnosis of many cardiac pathologies is rooted in gleaning patterns from \textit{synchronous} 12-lead ECG. If \textit{synchronous} 12-lead signals of clinical quality can be synthesized from these single-lead devices, it can transform cardiac care by substantially reducing the costs and enhancing access to cardiac diagnostics. However, prior attempts to synthesize \textit{synchronous} 12-lead ECG have not been successful. Vectorcardiography (VCG) analysis suggests that cardiac axis synthesized from earlier attempts deviates significantly from that estimated from 12-lead and/or Frank lead measurements. This work is perhaps the first successful attempt to synthesize clinically equivalent \textit{synchronous} 12-lead ECG from single-lead ECG. Our method employs a random forest machine learning model that uses a subject's historical 12-lead recordings to estimate the morphology including the actual timing of various ECG events (relative to the measured single-lead ECG) for all 11 missing leads of the subject. Our method was validated on two benchmark datasets as well as paper ECG and AliveCor-Kardia data obtained from the Heart, Artery, and Vein Center of Fresno, California. Results suggest that this approach can synthesize \textit{synchronous} ECG with accuracies ($R^2$) exceeding 90\%. Accurate synthesis of 12-lead ECG from a single-lead device can ultimately enable its wider application and improved point-of-care (POC) diagnostics.
\end{abstract}


\section{Introduction}\label{intro}
Cardiac disorders remain the leading cause of mortality claiming more than 17.3 million lives annually \cite{go2014heart}. An estimated 85.6 million people in the U.S. are living with cardiovascular diseases (CVD). Besides the elevated risk of mortality, CVD populations are confined to a much-degraded quality of life \cite{juenger2002health}. From a clinical standpoint, ECG tests remain the most common first step in the diagnosis of CVD. \textit{Synchronous} 12-lead ECG is used to provide a non-invasive and a fairly definitive diagnosis of cardiac disorders. The strategic electrode placement, costly and bulky ECG equipment, and a trained personnel for diagnosis requires the patients to visit a primary care center. Since about 42.2 million CVD population in the U.S. constitutes of elderly people ($\geq60$ years of age) with several comorbidities \cite{go2012heart,national2017health}, frequent visits to the primary care can be a challenging task. Untimely diagnosis usually results in significant myocardial damage, reduced survival rate, and even mortality \cite{benjamin2017heart}. Not to mention that a fifth of the heart attacks are silent where the person becomes aware of the symptoms weeks or months later. Hence, for patients under higher risk, a timely and often a continuous diagnosis is required.

Recent advances in remote health, POC, and wearable technologies have allowed for continuous/ day-to-day remote monitoring. The electronic health data thus generated provides a novel opportunity for developing a data-driven, personalized decision support system for the healthcare providers. Further facilitating timely diagnostics to enhance the quality of life and reduce mortality risks among CVD populations. One such technology is single-lead handheld/wearable ECG devices \cite{AliveCor.com,theheartcheck.com, irhythmtech.com}. These devices cost substantially less compared to the current clinical ECG recorders. They have significantly simplified recording ECG with a much easier application, access, and feasibility of everyday use. Consequently, they have caught the imagination of the medical community for transforming the clinical diagnostic practice. One such device, which we use in the current work, is the Alivecor-Kardia Mobile. It can gather ECG at 300Hz sampling rates, one-channel-at-a-time. These single-lead devices are beginning to be considered for detecting pathologies such as atrial fibrillation with promising initial results~\cite{orchard2016screening}.

However, a major technological barrier hampers the applicability of the current POC ECG technologies for clinical diagnostic applications--- medical practitioners are trained to parse patterns from 12-lead ECG signals to diagnose by correlating information from several leads. In fact, diagnosis of various forms of myocardial infarction and other acute cardiac conditions requires parsing patterns across two or more leads of \textit{synchronously} acquired ECG~\cite{wagner2009aha} and/or through the use of vectorcardiogram (VCG)~\cite{yang2012identification, yang2012spatiotemporal, le2013topology}.

A significant amount of work has been done for reconstructing 12-lead ECG from fewer leads. These efforts include reconstructing the missing leads either from a different lead system such as VCG or EASI~\cite{dower1988deriving,dawson2009linear} or from a reduced subset of standard ECG~\cite{tsouri2014patient}. All of these approaches require at least two  \textit{synchronously} acquired leads to reconstruct the 12-lead ECG. Hence they are not applicable to the present context. 

\begin{figure}[!htb]
\centering
		\includegraphics[width = 0.7\textwidth]{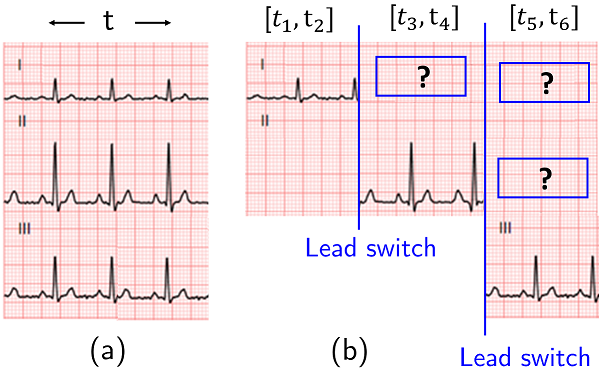}
		\caption{(a) Synchronous recording of ECG leads I, II, and III in clinical setting (b) recording lead I by placing the handheld device between a left and right arm or fingers during the time interval $[t1,t2]$. Then the device setting is switched to make the right arm and the left are touch the electrodes to record lead II signals during $[t2,t3]$, and subsequently switched to left arm-right leg during $[t3,t4]$ to record lead III signals. At most one lead signal is recorded at any given time.}
		\label{leadswitch}
\end{figure}

Attempts have been made to derive 12-lead ECG from single-lead portable devices~\cite{huang2016synthesizing}. These works are based on recording signals \textit{sequentially} from the single-lead device, one lead at a time. Electrode placement configurations have been developed for recording every lead signal from these devices~\cite{1leadECG}. For example, Fig.~\ref{leadswitch} (a) shows a snippet of the simultaneous recording of three ECG signals, namely lead I, II, and III using a traditional ECG machine. In contrast, Fig.~\ref{leadswitch} (b) shows an example of an asynchronous ECG recording with a single-lead device. A preliminary effort has also been made to derive 12-lead ECG from single-lead by a weighted linear combination of asynchronous bipolar signals. A recent pilot study used sequential 12-lead ECG recording from AliveCor\textregistered~heart monitor recorded by the trained research team for ST-elevation myocardial infarction (STEMI) diagnosis~\cite{muhlestein2015smartphone}. They demonstrated a potential for evaluation of acute ischemia using a single-lead device.

\begin{figure}[!htb]
\centering
		\includegraphics[width = 0.7\textwidth]{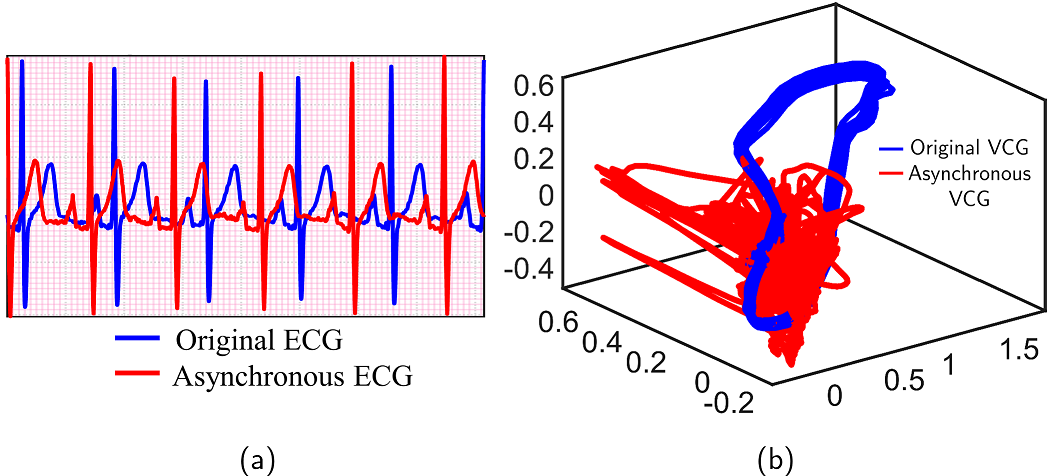}
		\caption{Comparison of ECG and VCG signals from measured (blue) signals versus those reconstructed from asynchronous (red) signals showing (a) significant differences in the timings of various ECG events of lead I, such as R peaks between the measured and the reconstructed signals. Note that the difference changes from beat to beat. (b) significant difference in the orientation of the QRS loop, and more than 90\textdegree~deviation in the estimation of the cardiac axis.}
		\label{vcg}
\end{figure}

However, the 12-lead signals derived from these earlier approaches are \textit{asynchronous}. Their VCG analysis suggests that cardiac axis synthesized from these earlier attempts deviates considerably from that estimated from 12-lead and/or Frank lead measurements. Fig.~\ref{vcg} shows an example of reconstructed lead I ECG and 3D VCG derived using the inverse Dower transform~\cite{dower1980deriving} of ECG signals recorded \textit{asynchronously}. As is evident from this figure, the VCG reconstructed from asynchronous signals deviates by more than {90}\textdegree~ from the measured signals. Consequently, every diagnostic method that employs VCG~\cite{le2013topology}, as well as diagnostic algorithms that use timing across multiple leads would be ineffective and can lead to misdiagnosis of critical CVDs.

Hence, an automated and accurate \textit{synchronous} reconstruction of the 12-lead ECG from a single-lead ECG device can be a game-changer in cardiac diagnostics and personalized healthcare, especially to substitute expensive instruments with more affordable wearable devices. To our knowledge, none of the earlier methods can synthesize \textit{synchronous} 12-lead ECG from a single-lead portable or wearable ECG device.

This paper presents a fully automated approach to derive \textit{synchronous} 12-lead ECG from single-lead devices with accuracies exceeding 90\% (in terms of $R^2$). The key innovation here is in our method to synchronize various lead signals by predicting the \textit{inter-lead temporal lags}, i.e., the difference in the timings of various ECG events across multiple leads based on a non-parametric machine learning model. The present approach consists of two stages: The first stage consists of obtaining the characteristic morphological and temporal information of the missing lead. The second stage consists of capturing the dynamic evolution in terms of inter-lead temporal lags and amplitude variations in the lead being currently recorded and incorporating it in the missing lead synthesis. Eventually, by doing so the entire matrix of missing leads can be completed to derive a 12-lead standard ECG. The remainder of this paper is organized as follows: the methodological details of synthesizing 12-lead standard ECG from single-lead ECG signal are presented in Section~\ref{method}. Section~\ref{data} presents details for the four datasets used in this paper to validate the efficacy of the proposed synthesis methodology. Subsequently, Section~\ref{results} presents the synthesis results for these datasets. Lastly, we conclude with discussions in Section~\ref{discusssion}.
 
\section{Methodology}\label{method}
Synthesizing a missing ECG lead essentially involves estimating three key aspects of the signal: (a) morphology of the waveform, (b) temporal information or timing of the fiducial points, and (c) the dynamic evolution of morphological and temporal features in terms of inter-lead temporal lags and amplitude variations. Morphology refers to the shape characteristics of an ECG wave, such as T-wave pattern, its skewness, elevation, etc., and the amplitudes of the various waves of a signal in every beat. Accurate morphological information across different leads is necessary to diagnose acute conditions such as arrhythmia. Temporal information refers to the timestamps of events, such as the onset, offset, and peaks of various waves such as p-wave, T-wave etc. within every beat of a signal. Accurate temporal information is necessary to estimate the lengths of various intervals, such as QT and RR, which are employed to diagnose acute conditions such as STEMI and to track the progression of a CVD. The morphology information for any missing lead at time $t$ is obtained from their corresponding historic lead recordings. 

\begin{figure}[!htb]
\centering
		\includegraphics[width = .9\textwidth]{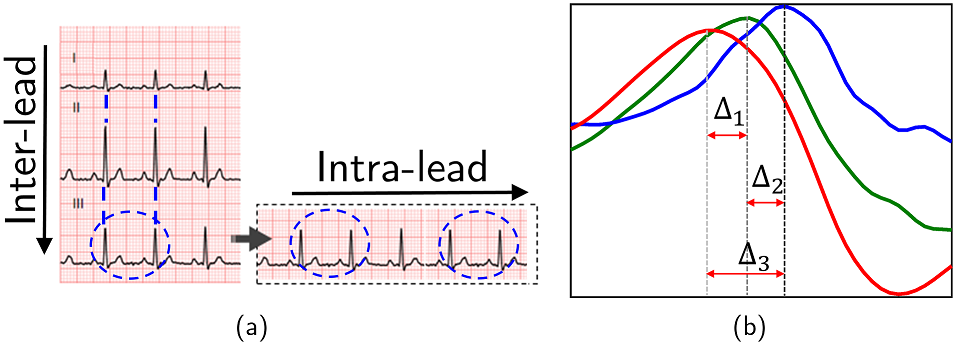}
		\caption{(a) Intra-lead morphology and inter-lead temporal alignment in the ECG signal that forms the basis for the primary synthesis of missing lead (b) representation of the inter-lead temporal lags in the R-peak of lead I, II, and III of a clinical ECG.}
		\label{current}
\end{figure}

More precisely, when a subject is prescribed a handheld ECG system, the healthcare professional can upload their simultaneous 12-lead ECG recording (recorded most easily with any traditional ECG machine) to a cloud-based server. We refer to this simultaneous 12-lead ECG as ``historical data". Further, we know that at any time $t$, the handheld device is used to record a single-lead only (referred to as the ``current lead" from now on). Thus, the temporal information is obtained from the signal being currently recorded at time $t$. The rationale for doing so lies in the quasi-periodic nature of the heart rhythms, as a consequence of which, the overall morphology information, i.e., fiducial shape holds significant similarity within a lead (intra-lead morphological correlation). However, even within a lead, there are dynamic variations in both horizontal scale, i.e., timing and vertical scale, i.e., voltage. Nonetheless, these dynamic variations are highly correlated between synchronously recorded leads (inter-lead temporal correlation). Fig.~\ref{current}~(a) presents a segment of clinical 12-lead ECG showing this intra-lead morphology and inter-lead temporal alignments.

While this morphological and temporal information is sufficient for an initial synthesis of the missing leads, the synthesized signals obtained thus are far from accurate due to the error in temporal alignment as a result of inter-lead temporal lags. The inter-lead temporal lags refer to the differences in the time stamp of an event, e.g., the timing of R peak, as recorded across various leads for the same beat. From a physiological standpoint, these lags emerge because each lead signal is a different projection of the spatio-temporal evolution of the electrocardiac activity~\cite{le2013topology}. Consequently, the timings of the peak values and other events would be different across the various leads. Fig.~\ref{current}~(b) shows the inter-lead temporal lags in the R peak of lead I, II, and III of a clinical 12-lead ECG recording.

Since the event-timing information of the current lead is used for missing beats synthesis, unresolved inter-lead temporal lag results in bias (as illustrated in Fig.~\ref{vcg}) in the timing of reconstructed signals. However, we are unaware of any work in the missing lead synthesis/reconstruction literature that has brought out this aspect.

In this method, once a beat using the single current lead is generated during the stage I, the timings of this synthesized beat are adjusted with the lag values for various events, learned using a machine learning procedure. Using just a single-lead for \textit{synchronous} reconstruction of missing leads and the lag correction forms the key innovations of the present method. In the next two subsections, we describe the steps for synthesizing the missing leads using the morphology and temporal information from the historic and current leads, respectively, prediction of inter-lead lag, and obtaining the final synthesis after lag correction.

\subsection{Obtaining the morphology and temporal information of the missing leads}\label{method1} 
For the proposed method, extracting the morphological and temporal information happens on a beat to beat basis (RR interval). Hence, the first step after signal pre-processing (noise and baseline wandering removal) is to obtain the fiducial point locations. We use a combination of the Pan-Tompkins and QRS detection algorithm to correctly obtain the fiducial point locations in the current and historical lead signals~\cite{engelse1979single,pan1985real,laguna1994automatic}. Along with robust determination of the fiducial point, the algorithm adapts to the different morphologies of the ECG signal which is crucial for accurate reconstruction in signals with T-wave alternans (TWA) and other abnormalities such as biphasic and inverted T-waves (accuracy shown in the result section).

\begin{figure}[!htb]
\centering
		\includegraphics[width = .7\textwidth]{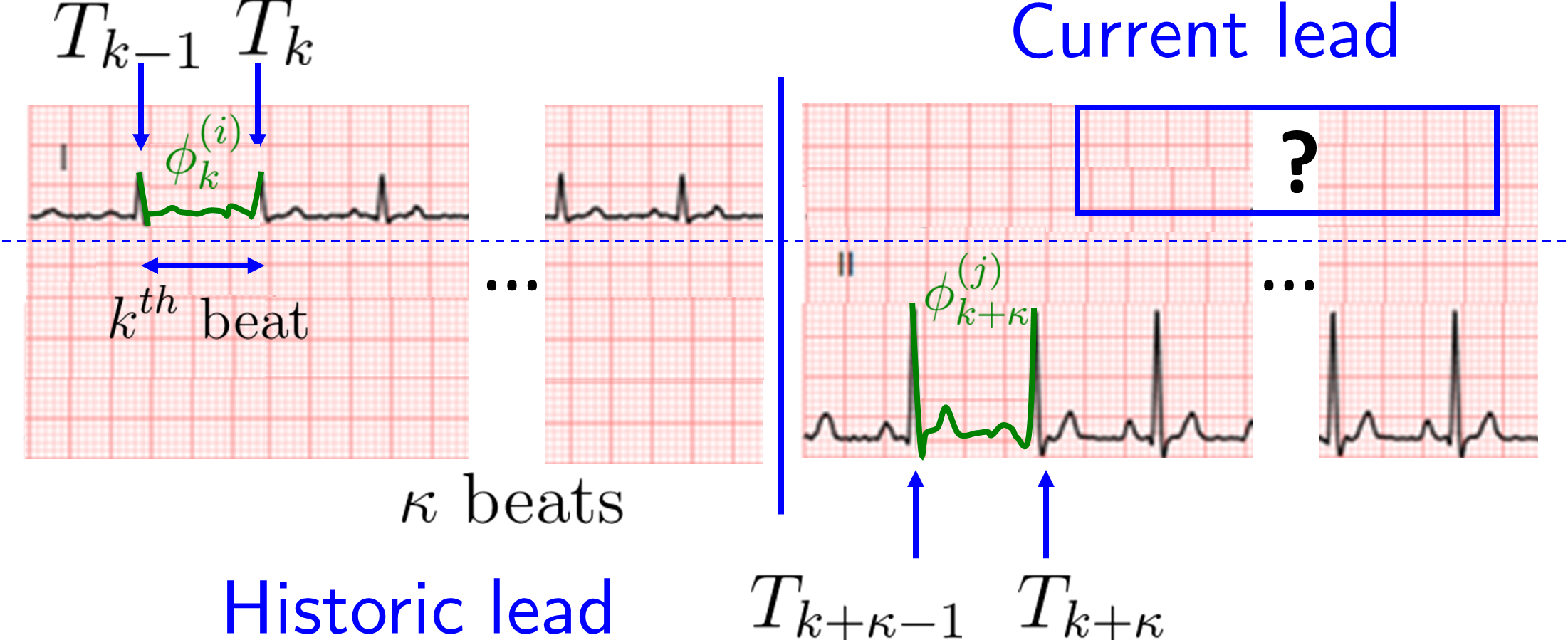}
		\caption{Every historic and current lead is divided into segments, $\phi_k$, each spanning an RR interval $[T_{k-1}, T_k]$ between two successive beats. Morphological information gathered from the historic lead and temporal information gathered from the current lead are used to reconstruct a missing lead.}
		\label{formulation}
\end{figure}

Let $\phi_k^{(i)}(t|T_{k-1}^{(i)}\leq t \leq T_{k}^{(i)})_{\left\{1\leq i\leq 12\right\}}$ represent $k^{th} (k\geq 1)$ RR interval of lead $i$. Let the length of this beat be ${\delta}_{k}^{(i)}={T}_{k}^{(i)}- {T}_{k-1}^{(i)}$, with the first R peak of the $k^{th}$ RR occurring at time $T_{k-1}^{(i)}$ and the second occurring at time $T_{k}^{(i)}$. Finally, let there be $\kappa>0$ beats recorded before the handheld device is switched to a different position and recording for another lead, $j$ starts. For example, as shown in Fig.~\ref{formulation}, after recording $\kappa$ beats, the single-lead device is switched from one location to another. Specifically this figure shows $i$ to be lead I and $j$ to be lead II. Now, when the ${k+\kappa}^{th}$ RR interval (ignoring the missed beats during lead switch) for lead $j$ is recorded, there is no corresponding recording for lead $i$. The synthesis of the ${k+\kappa}^{th}$ RR interval for lead $i$, $\tilde{\phi}_{k+\kappa}^{(i)}$ can be given as: 

\begin{equation}
%\begin{split}
\tilde{\phi}_{k+\kappa}^{(i)}(t|\tilde{T}_{k+\kappa-1}^{(i)}\leq t \leq \tilde{T}_{k+\kappa}^{(i)})=\pazocal{T}(\phi_{k+\kappa}^{(ij)}(t|T_{k+\kappa-1}^{(ij)} \leq t\leq T_{k+\kappa}^{(ij)}))
%\end{split}
\label{RR}
\end{equation}

\noindent where, $\pazocal{T}$ represents affine transformation of the historic RR interval of lead $i$ that has maximum similarity to the current ${k+\kappa}^{th}$ RR interval of lead $j$, $\phi_{k+\kappa}^{(ij)}$. Here we used the similarity between RR intervals of lead $i$ and $j$ is measured using dynamic time warping \cite{muller2007dynamic}, which allows comparing time-dependent RR intervals with different lengths. This transformation in (\ref{RR}) consists of two essential steps: (i) the energy (voltage) of the synthesized beat is matched with the current beat's energy and (ii) due to the temporal correlation explained above, the time duration for the synthesized RR interval of lead $i$ is considered synchronized with the RR time duration of the current lead $j$, i.e., $\tilde{\delta}_{k+\kappa}^{(i)} = \delta_{k+\kappa}^{(j)}$. Once each RR interval of the missing lead is synthesized, they are concatenated to obtain a complete initial reconstruction of the missing lead.  

\begin{figure}[!htb]
	\centering
	\includegraphics[width = .7\textwidth]{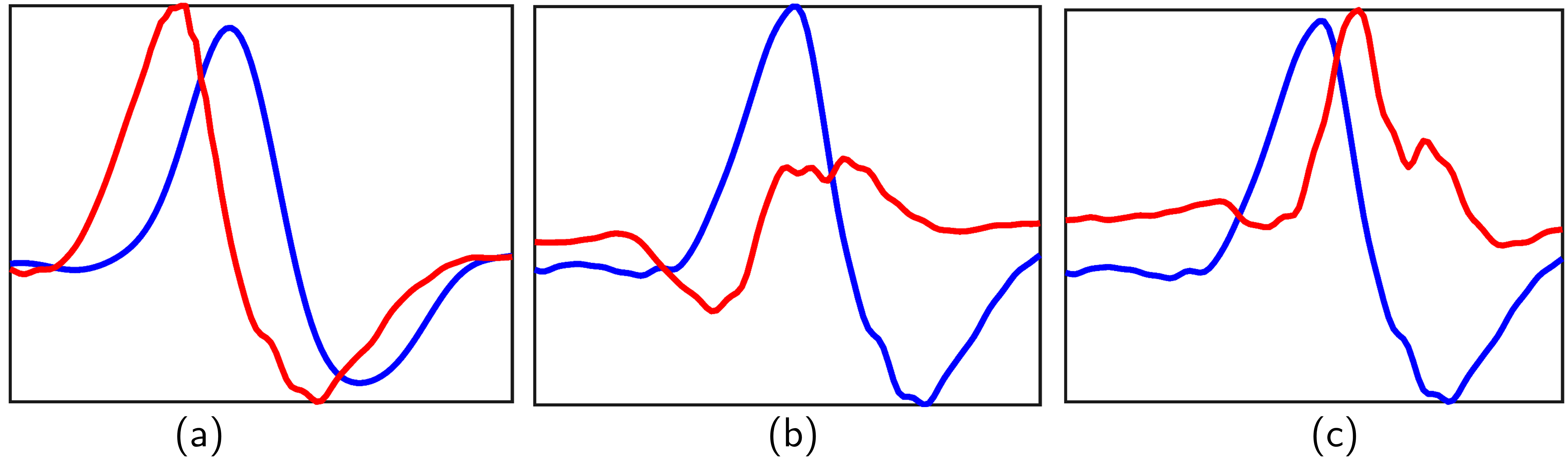}
	\caption{Representation of the temporal bias/misalignment as a result of the inter-lead temporal lag between the actual (blue) and synthesized (red) R beat (a) using the proposed method without lag correction, (b) a recently proposed method of weighted linear combination of asynchronous bipolar leads obtained from a handheld ECG device~\cite{huang2016synthesizing}, and (c) a weighted linear combination of \textit{synchronous} bipolar leads obtained using traditional ECG machines (similar method~\cite{nelwan2000minimal}).}
	\label{gap1}
\end{figure}

However, due to the inter-lead event lags, considering $\tilde{\delta}_{k+\kappa}^{(i)} = \delta_{k+\kappa}^{(j)}$ results in significant reconstruction error between the synthesized and the original measured signals (as shown in Fig.~\ref{gap1}). Fig.~\ref{gap1} (a) shows this error in reconstructed missing beat using the proposed method without correcting the temporal lag. In order to compare how other methods perform in presence of the temporal lag, we used the methodology proposed in one of the recent work in ECG reconstruction using single-lead \cite{huang2016synthesizing}. The missing beat was reconstructed using a weighted linear combination of asynchronous bipolar leads (we used equal weights) and its reconstruction accuracy is shown in Fig.~\ref{gap1} (b). Given that weighted linear combination is a popular method in the ECG reconstruction literature~\cite{finlay2007synthesising}, we linearly combined \textit{synchronous} bipolar leads using equal weights to compare the reconstruction accuracy. The reconstructed beat as compared to the original beat is shown in Fig.~\ref{gap1} (c). 

Since, this inter-lead event lag is an inherent characteristic of ECG leads, irrespective of the reconstruction method used, it always resulted in a temporal bias. Hence, after obtaining the primary reconstruction we employ the lag correction method (detailed in the next subsection) to obtain the final reconstructed signal.

\subsection{Inter-lead lag prediction and correction }\label{RF}
As detailed in the foregoing, there is an inherent temporal lag even between the synchronously recorded leads of ECG. This lag essentially depends on the person-to-person, lead-to-lead, as well as beat-to-beat changes in the spatiotemporal distribution of the cardiac electrical activity. Further, due to the heart rate variability, the RR interval is time-varying, thus making it a non-trivial task to model this lag. In this work, we used random forests (RF) regression model to predict this dynamic and non-linearly varying inter-lead event lag. Along with improved robustness and generalization, the performance of RF for various applications is comparable to most state-of-the-art methods \cite{strobl2009introduction,arumugam2016random, afrin2018balanced}. As mentioned above, the lag depends on person-to-person and from lead-to-lead, hence the main objective here is to use the historical data in order to build a personalized lag prediction model.

RF \cite{breiman2001random} is an ensemble of decision trees generated using B bootstrapped samples from the training data, $\mathcal{D}=\left\{\left(\bm{x}_l, \Delta^{(ij)}_l\right) \right\}_{\left\{1\leq l\leq N\right\}}$, where $\bm{x}_l={x^{(j)}_1, ..., x^{(j)}_P}$ is a $P$ dimensional feature vector extracted from the historical data corresponding to the current lead $j$ (assuming lead $i$ is missing). $\Delta^{(ij)}_l$ is the inter-lead lag between the RR interval lengths of any missing lead, $i$ and current lead, $j$, i.e., $\Delta_{l}^{(ij)} = {\delta}_{l}^{(i)}-\delta_{l}^{(j)}$, and $N$ is total number of training data samples. Since during the prediction, only the signal from current lead is available, the training data also constitutes features extracted only from the historic recording corresponding to the current lead only.

More precisely, the construction of $b^{th}$ bootstrapped decision tree, $\mathcal{T}_b$ begins by randomly selecting $m$ out of $P$ predictors and recursively splitting the predictor space so as to minimize the prediction error of each region thus formed. The value of $m$ if usually taken as the square root of the total predictor size, i.e., $m=\sqrt{P}$ \cite{james2013introduction}. The recursive split continues until a termination criterion is not reached (which in most case is the minimum number of observations in the terminal node). Finally, the prediction error is the average of the prediction error for all the trees in the ensemble. This aggregated results of the ensemble provide a more robust result than a single decision tree. 

During the training phase of the RF regression model for inter-lead lag prediction, the response vector is the inter-lead lag, $\Delta^{(ij)}$ between the historical recordings of any missing lead $i$ and the current lead $j$. However, the input vector, $\bm{x}$ contains the features extracted just from the current lead's data. This is done because during the testing, only data from current lead will be available. We utilized 60 seconds of data to obtain the predictors and response for training the model. The fiducial point locations including their onset and offset time in every RR interval. The input feature vector, $\bm{x}$ then constitutes the temporal and amplitude features of all fiducial points (as shown in \autoref{featuretable}) extracted for each RR interval of the current lead's data. Among the 10 extracted features, R height, RR interval, T height, and ST duration were the most significant.

\begin{table}[!htb]\centering
	\renewcommand{\arraystretch}{1.3}
\caption{Predictors used in the RF regression model. For training the model, these predictors are extracted for every beat of the current lead's historical data and for testing they are extracted from every beat of the current lead.}
\label{featuretable} 
\setlength{\tabcolsep}{3pt}
\begin{tabular}{c c}
\hline
\textbf{Amplitude Features} & \textbf{Temporal Features}\\
\hline
	R height ($\mu V$) & RR interval (ms)\\
			P height ($\mu V$)& QRS duration (ms)\\
			T height ($\mu V$) & ST duration (ms)\\
			S peak ($\mu V$)& PR duration (ms) \\
			Q peak ($\mu V$) & T-wave duration (ms) \\
\hline
		\end{tabular}
\label{feature}
\end{table}

\begin{figure}[!htb]
\centering
		\includegraphics[width = .7\textwidth]{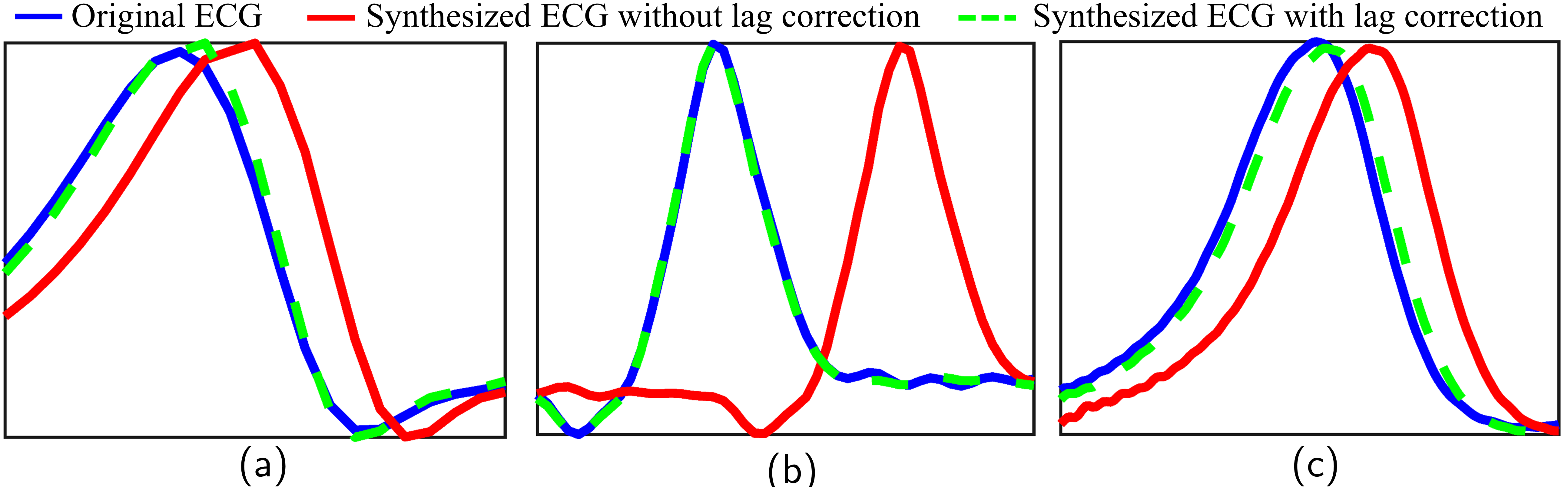}
		\caption{Performance of missing beat synthesis before (in red) and after (in green) the inter-lead temporal lag correction using random forest algorithm for a representative subject. The subfigure represents the synthesized signal for missing (a) lead III when V3 is the current lead (b) lead V6 when III is the current lead, and (c) lead V4 when V1 is the current lead. }
		\label{gap2}
\end{figure}

Once the predicted lag is obtained, the RR interval length of the synthesized lead is corrected, i.e, $\tilde{\delta}_{k+\kappa}^{(i)}=\delta_{k+\kappa}^{(j)}+\Delta_{k+\kappa}^{(ij)}$. This constitutes the final key step. Fig.~\ref{gap2} shows the original ECG signal/ground truth and the synthesized signal with and without the lag correction methodology. This figure shows the temporal misalignment between the actual signals and the signal synthesized without lag correction. In contrast, the synthesized signal obtained after the lag correction using RF regression has a significantly improved alignment with the actual signal. Now, once every RR interval is corrected for the lag error, they are concatenated together to give the final synthesized lead. In this work, we used only $60$ seconds of data for training the RF regression model, a larger training dataset can further improve the lag prediction accuracy. 

The main steps of the lead synthesis methodology presented in this paper are summarized in Algorithm~\ref{alg}. Missing lead synthesis accuracy and the datasets used for validation are detailed in the following sections. 

\begin{algorithm}[!htb]
		\begin{algorithmic}[1]
			%\STATE Input:  $i\leftarrow 1, b \leftarrow 1, x^* \leftarrow0, \zeta^* \leftarrow 0$
			\STATE Input \& Preprocess: lead $i\leftarrow$ historic lead, lead $j\leftarrow$ current lead
			\STATE Extract: Fiducial Points- Pons, Ppeaks, Poffs, Qpeaks, Rpeaks, Speaks, Tons, Tpeaks, Toffs for lead $i$ and $j$
			\STATE Extract: Amplitude and temporal features for lead $j$ {\color{red}\%\textit{see Table~\ref{feature}}}
			\STATE $n \leftarrow$ number of RR intervals in lead $j$
			\FOR{ $k=1:n$}
			\STATE Obtain: $\phi_{k+\kappa}^{j}$,  $\phi_{k+\kappa}^{ij}$,  $\delta_{k+\kappa}^{j}$
			\STATE Assign morphology: $\tilde{\phi}_{k+\kappa}^{i}=\phi_{k+\kappa}^{ij}$
			\STATE Predict: $\Delta_{k+\kappa}^{ij}$ using RF algorithm  {\color{red}\%\textit{see section~\ref{RF}}}
			\STATE Assign temporal: $\tilde{\delta}_{k+\kappa}^{i}=\tilde{\delta}_{k+\kappa}^{ij}+\Delta_{k+\kappa}^{ij}$
			\ENDFOR
			\FORALL{${k} \in \left\{1,n\right\}$}
			\STATE Concatenate $\tilde{\phi}_{k+\kappa}^{i}$
			\ENDFOR
		\end{algorithmic}
		\caption{Missing lead synthesis}
		\label{alg}
	\end{algorithm}

\section{Data and Experimental Protocol}\label{data}
We validated the methodology on two benchmark datasets from Physionet \cite{goldberger2000physiobank}. The first set was the PTB database \cite{bousseljot1995nutzung}. The database consists of 80 recordings acquired from 54 healthy volunteers and 368 MI recordings from 148 patients. Each recording in the PTB database contains 15 synchronously recorded signals, namely, the conventional 12-lead ECGs and the three orthogonal Frank XYZ leads. Signals were sampled at 1 kHz rate, with a 16-bit resolution over $\pm 16.384$ mV range. The second set of data from Physionet was the T-wave alternans (TWA) database \cite{moody2006physionet}. This database consists of a multichannel recording of 100 subjects with varying degree of TWA. The database includes patients with sudden cardiac death risk factors such as myocardial infarction, transient ischemia etc. The ECG signals were sampled at 500 Hz with 16-bit resolution and $\pm 32$ mV range. Although for some subjects only 2-3 lead ECG recordings were present, we used the data for subjects with all 12 synchronous recording. Further, subjects with a higher level of TWA were chosen. 

For the handheld device, we used the data from Alivecor-Kardia recorded at Heart, artery, and Vein center of Fresno. This dataset consisted of 101 ECG recordings of Kardia device from AliveCor\textregistered~taken from healthy as well as subjects with acute CVDs. A clinical staff member gathered multiple ECG signals following the 12-lead ECG extraction protocol specified for the Kardia device and saved as PDF files (note that the signals are asynchronous). Signals were recorded at a sampling frequency of 300 Hz and 16 bit A/D resolution. Thereafter, paper ECG signals were collected from each of the subjects using a Burdick-Mortara machine from each of the subjects. The paper ECGs were scanned and subsequently digitized using im2graph software (v.1.20). AliveCor\textregistered~PDF files were directly used for digitization using the im2graph software. The signals were preprocessed to remove scanning-induced distortions and make the data file compatible with the Physionet database. As a final step, the fiducial beats were extracted as described in section~\ref{method1}. The datasets from the first two sources were employed for direct validation and benchmarking of the performance of the proposed approach. However, due to the unavailability of ground truth, Kardia and paper ECG datasets were essentially employed for visual comparison of the morphological and temporal patterns of the reconstructed ECG.
  
\section{Results}\label{results}
\begin{figure}[!htb]
\centering
		\includegraphics[width =.7\textwidth]{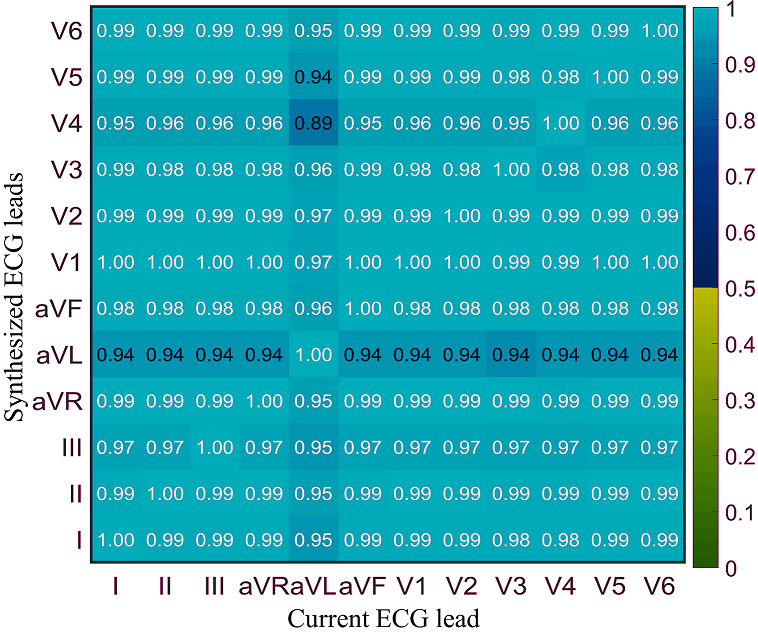}
		\caption{Accuracy matrix for the synthesis of missing 11 leads given any one of the 12 leads was recorded using the handheld device. Correlation coefficient accuracies are shown for a representative healthy subject in the PTB database with an average accuracy of $96\%$}
		\label{acc}
\end{figure}

In this section, we present the results for synthesizing the missing lead signals for the ECG datasets described in section~\ref{data}. For Physionet datasets with all 12-lead signals available, we assumed that the handheld ECG device can be used to record either of the 12-leads at a given time $t$ and for that time period rest of the 11 signals will be missing. For example, if lead I was used as the current lead for time duration $t$, we can synthesize the rest of the leads using the proposed methodology. Likewise, when the device is switched from lead I location to lead II location at time $t+\Delta t$, all the missing leads at that time can be synthesized and so on. Hence, one can consider this analogous to filling missing elements of a $12 \times 12$ matrix as shown in Fig.~\ref{acc}. In this figure, the column labeled ``I" shows the coefficient, $\rho$ accuracies values for synthesizing all other leads when lead I is the current lead. On the other hand, the row labeled ``I" in this figure represent the accuracies of reconstructing lead I using rest of the leads as the current lead. Naturally, the diagonal elements have a value of $1$. As is evident from this figure, the reconstruction accuracies are significantly high for every missing lead synthesis.

\begin{figure}[!htb]
\centering
		\includegraphics[width = .7\textwidth]{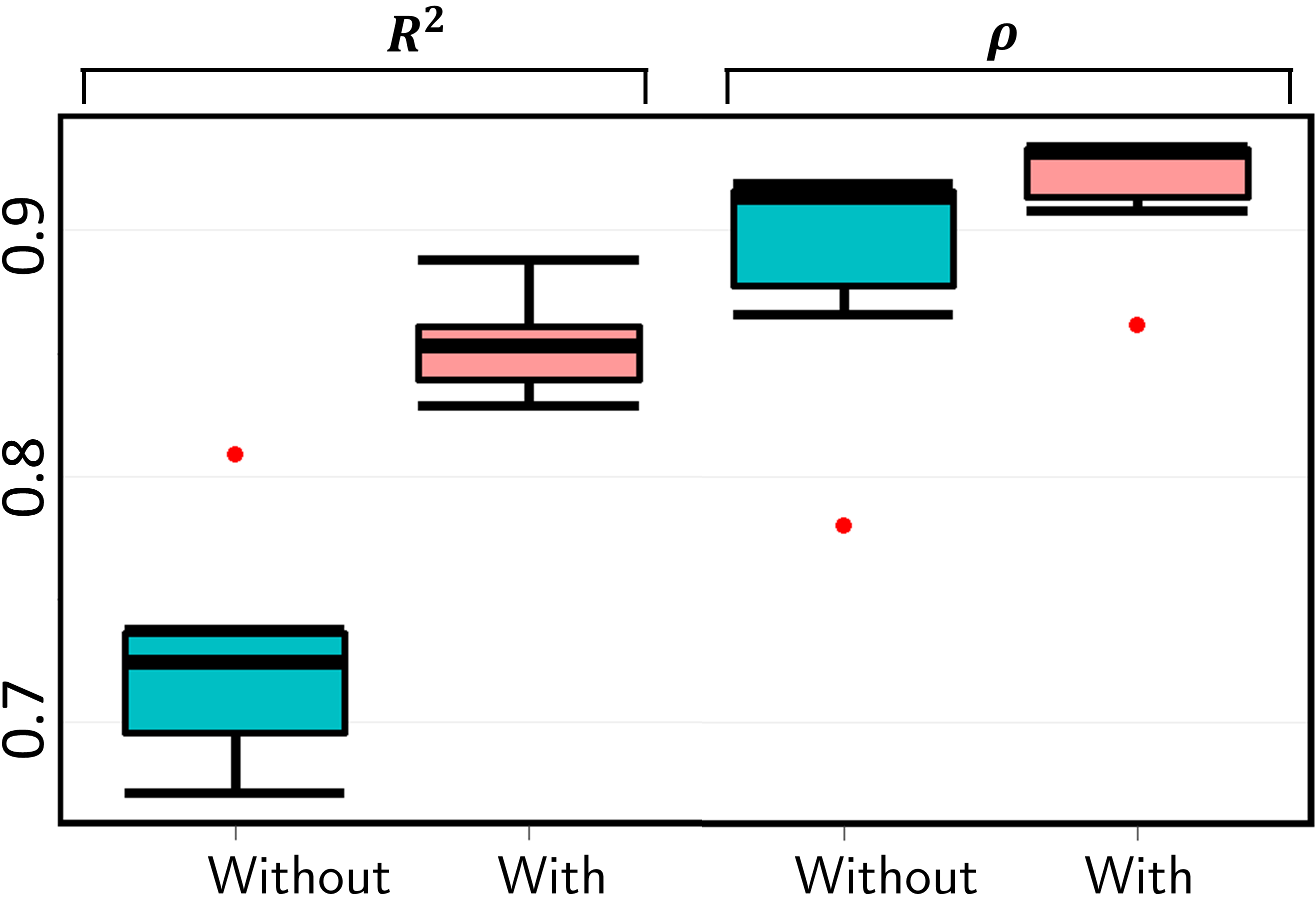}
		\caption{Boxplot demonstrating the accuracy improvement (in terms of $R^2$ and $\rho$) for reconstructing all missing leads using lead III of a subject in PTB-TWA database ``with” and ``without” the proposed inter-lead lag correction. The boxplot is bounded by the 25th and 75th percentile representing the inter-quartile (IQR) range. The middle horizontal line represents the median accuracy value.}
		\label{boxplot}
\end{figure}

Further, Table~\ref{accuracy} presents the average synthesis accuracies (across all $12 \times 12$ matrix) in terms of $R^2$ and $\rho$ values and their respective standard deviation. These accuracies are presented for two online datasets obtained from Physionet and are categorized for subjects with different health conditions. Due to the subtle physiological beat-to-beat variations in the TWA dataset, previous works were unable to reconstruct the signals \cite{langley2010estimation}. However, our method provides a reasonably high accuracy even for the TWA dataset. These high accuracy values are a result of inter-lead lag correction. Fig.~\ref{boxplot} summarizes the improvement in accuracy after the lag correction with blue boxplot representing the synthesis accuracies without the lag correction and the boxplots in red represents the synthesis accuracies with lag correction. 

\begin{table}[!htb]\centering
		\renewcommand{\arraystretch}{1.3}
\caption{Average $R^2$ and $\rho$ and their standard deviation (std) for the prediction across all the leads of representative cases of Physionet PTB and TWA databases. }
\label{accuracy}
\begin{tabular}{l  c  c }
\hline
		\textbf{Dataset} & \textbf{Average $R^2$ {\color{blue}(std)}}& \textbf{Average $\rho$ {\color{blue}(std)}} \\
\hline
		\textbf{PTB-Healthy}  & 0.91 {\color{blue}(0.06) }& 0.96 {\color{blue}(0.02) }\\ 
		\textbf{PTB-MI} & 0.95 {\color{blue}(0.03)} & 0.97{\color{blue}(0.02)}\\
		\textbf{TWA }& 0.93 {\color{blue}(0.02)} & 0.96 {\color{blue}(0.03)} \\
\hline
	\end{tabular}
\end{table}

Although Table~\ref{accuracy} shows high overall synthesis accuracy, we are interested in synthesizing missing signals using only the leads which can be recorded most conveniently using a handheld ECG device such as lead I and the precordial leads. Since lead I is provided by most ECG devices, we selected lead I as the current lead. We notice that (evident from Table.~\ref{accuracy_all}) lead I can very successfully be used to synthesize the rest of the missing leads. Additionally, using the precordial leads (which can also be recorded with ease) as current lead also resulted in a high accuracy of missing lead synthesis.

\begin{table}[!htb]\centering
\renewcommand{\arraystretch}{1.3}
\caption{Average accuracy for synthesizing missing 11 leads using lead I as the current lead for healthy and unhealthy cases in the PTB database}
\label{accuracy_all}
\begin{tabular}{l  c  c }
\hline
	\textbf{Subject} & \textbf{Average $R^2$ {\color{blue}(std)}} & \textbf{Average $\rho$ {\color{blue}(std)}}    \\
\hline
			\textbf{Healthy Cases} & 0.93 {\color{blue}(0.06)} & 0.97 {\color{blue}(0.03)} \\
			\textbf{Unhealthy Cases} & 0.92 {\color{blue}(0.06)} & 0.96 {\color{blue}(0.03)}\\
\hline
		\end{tabular}
\end{table}

\begin{figure}[!htb]
\centering
		\includegraphics[width = .7\textwidth]{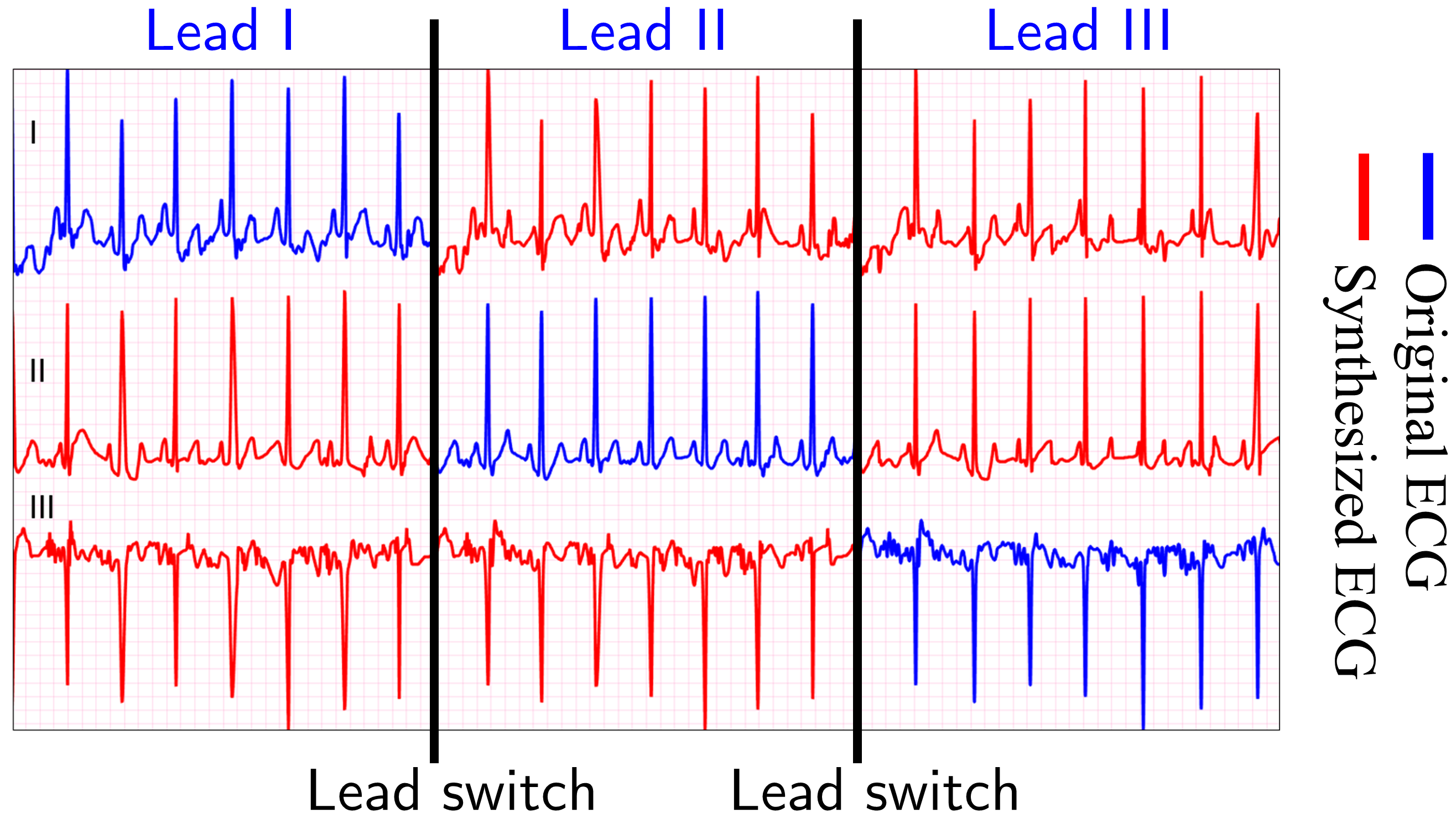}
		\caption{Obtaining asynchronous single-lead recordings (blue) for lead I, II, and III using the AliveCor-Kardia and synthesizing the missing leads (red). Whenever the handheld device was switched from one location to another, the current lead was used to synthesize the other two missing leads.}
		\label{kardia}
\end{figure}

\begin{figure}[!htb]
\centering
		\includegraphics[width = .7\textwidth]{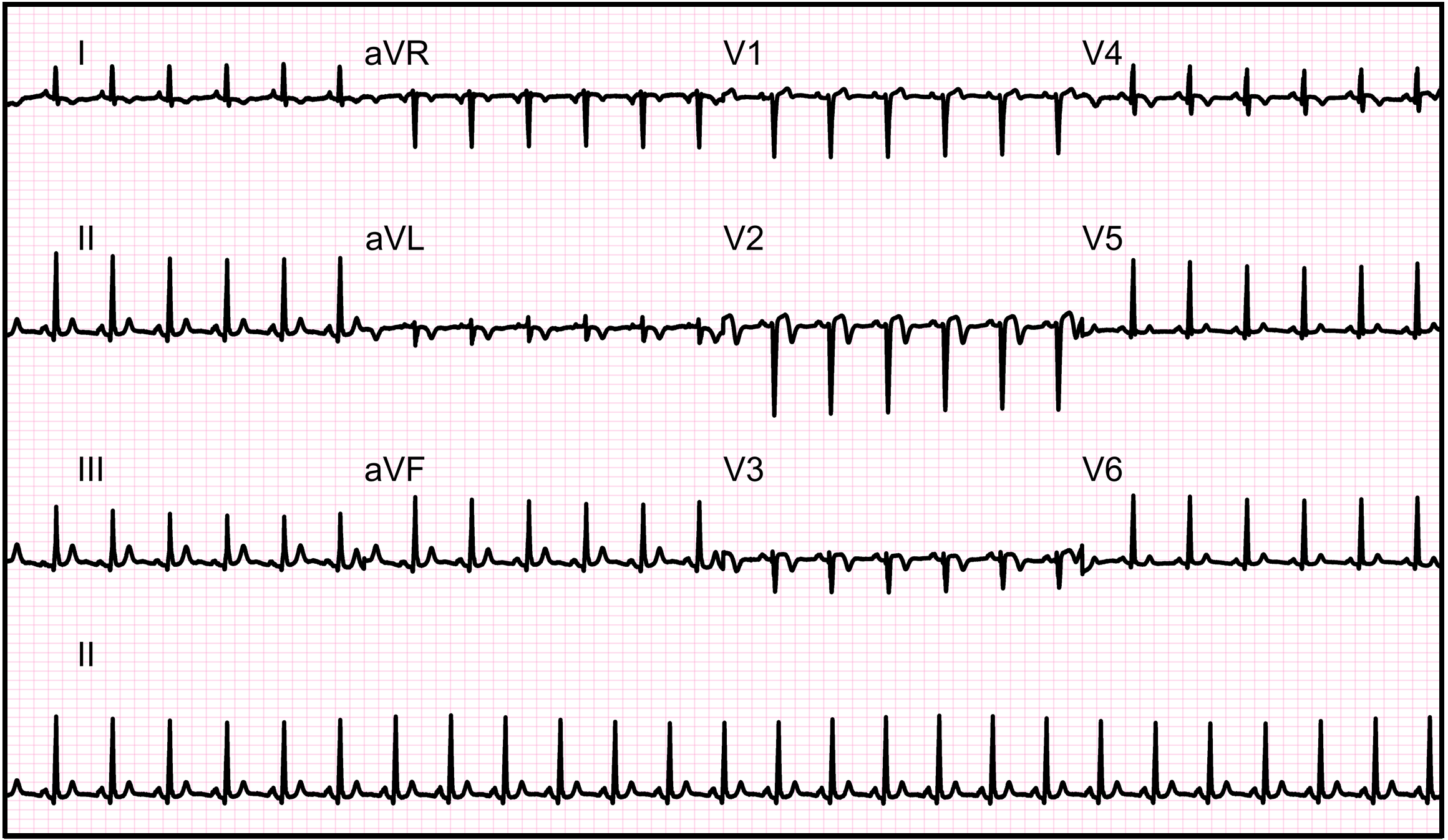}
		\caption{Representation of missing lead synthesis using the digitized paper ECG database. For an unhealthy subject, all missing ECG leads were synthesized assuming lead I as the current lead.}
	\label{ecgallpaper}
\end{figure}

Finally, Fig.~\ref{kardia} presents the synthesized lead signals for the AliveCor-Kardia device. Here, the signals shown in blue are the current signal and the ones shown in red are synthesized using the proposed method. A visual comparison between the actual and synthesized signal for all three leads corroborates the efficacy of the proposed method. Further, we applied the proposed method to the digitized paper ECG database. Only lead I was utilized to reconstruct all other missing leads (Fig.~\ref{ecgallpaper}). There was no visual difference between the actual and reconstructed signals. 

\begin{figure}[!htb]
\centering
		\includegraphics[width = .7\textwidth]{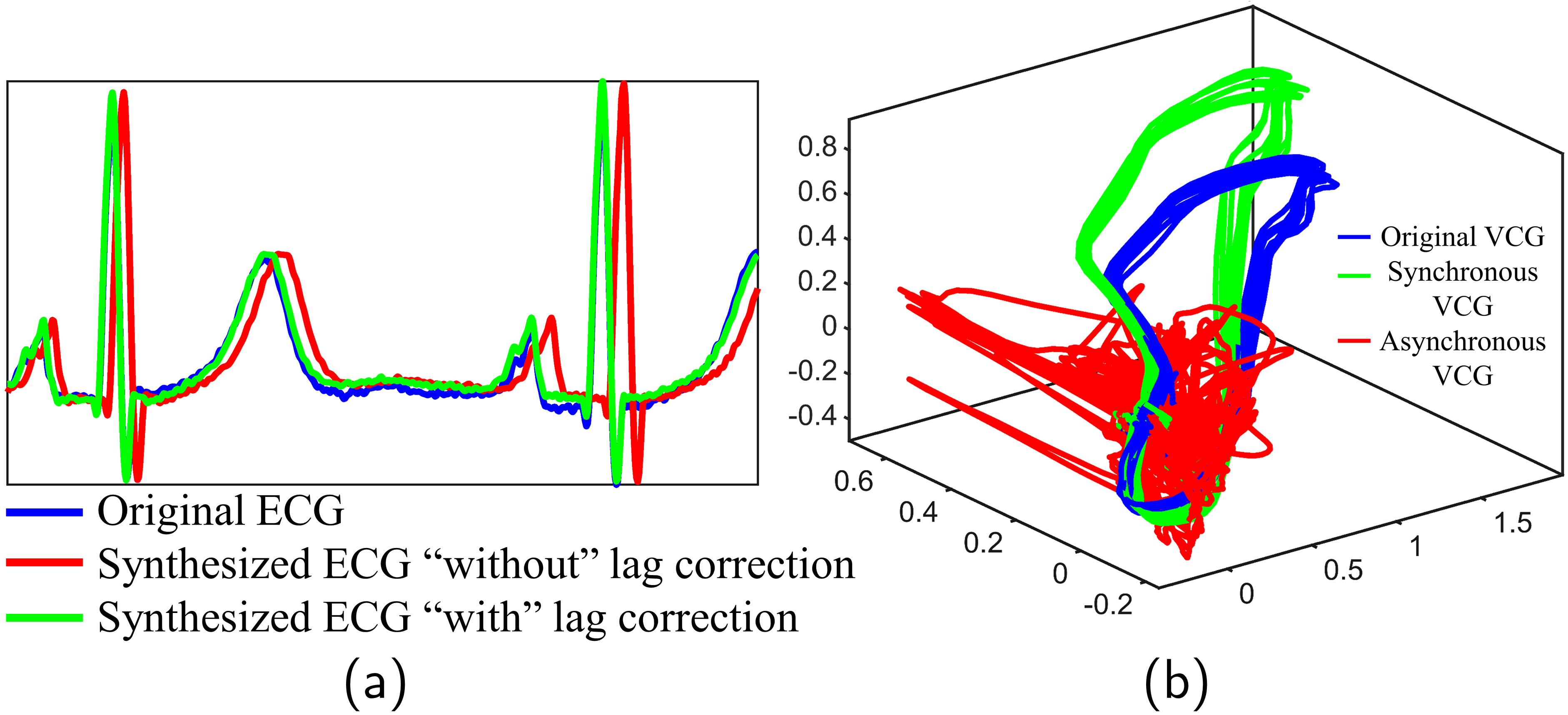}
		\caption{Comparison of measured ECG (blue) signals versus those reconstructed using the proposed method “with” and “without” gap correction for a representative subject from TWA database. (a) The accuracy of synthesizing lead III using lead I, $\rho= 0.41$ increased to $\rho = 0.97$ after gap correction (b) VCG thus reconstructed is significantly aligned with the original VCG obtained using inverse Dower transform.}
	\label{vcgsync}
\end{figure}

Finally, Fig.~\ref{vcgsync} demonstrates the main contribution of this work: (i) the accurate synchronous reconstruction of all the missing leads (and thus any derived signal such as VCG) using a single-lead handheld device (Fig.~\ref{vcgsync} (a)) and (ii) obtaining accurate dynamic temporal information by the inter-lead lag prediction methodology (Fig.~\ref{vcgsync} (b)). The reconstructed ECG, as well as VCG, has excellent alignment with their respective measured signals. To our knowledge, this study is the first to do an accurate and \textit{synchronous} reconstruction of missing leads using a single-lead ECG device.

\section{Discussion and Future Work}\label{discusssion}
In this paper, we proposed a potentially pioneering work for synthesizing \textit{synchronous} 12-lead ECG using single-lead handheld ECG devices. Since multiple signals recorded from the single-lead device are asynchronous, we showed that the traditional methods for 12-lead reconstruction from a subset of available leads yield highly unreliable results. Furthermore, we demonstrated the bias in accuracy arising due to the inter-lead lag, which has not been previously addressed in the extant literature. Finally, we developed a personalized analytics for the prediction of this lag and thereby improve the accuracy of synthesizing the missing signals. For a subject with TWA, the accuracy increased by around 50\% (in $R^2$) and 20\% (for $\rho$) after the gap correction. On an average across all the datasets, $R^2$ and $\rho$ increased by 25\% and 13\%, respectively.

From a scientific standpoint, conventional wisdom as propounded in Dower’s seminal work suggests that signals synchronously acquired from an equivalent of three orthogonal leads would be needed to reconstruct 12-lead ECG. This viewpoint has been challenged in the literature, with some arguments noting the necessity for more, and the others suggesting the sufficiency of fewer than three leads for accurate reconstruction. These efforts have opened the key question of, how well can one reconstruct 12-leads using the information gleaned from a single-lead.

Recent advances in nonparametric machine learning models, such as random forests, allow us to pull almost all information pertinent for 12-lead reconstruction from a single-lead signal without introducing undue biases in terms of model structure or preferences for the ECG features. The present work is one of the initial efforts in extracting this information for 12-lead reconstruction. The results from this effort strongly suggest that information necessary for accurate reconstruction 12 leads can be gleaned from any one of several leads. The foregoing results also demonstrate the potential to further enhance the accuracy of reconstruction by adapting the rapidly growing repertoire of advanced machine learning models 
\begin{figure}[!htb]
\centering
		\includegraphics[width = .7\textwidth]{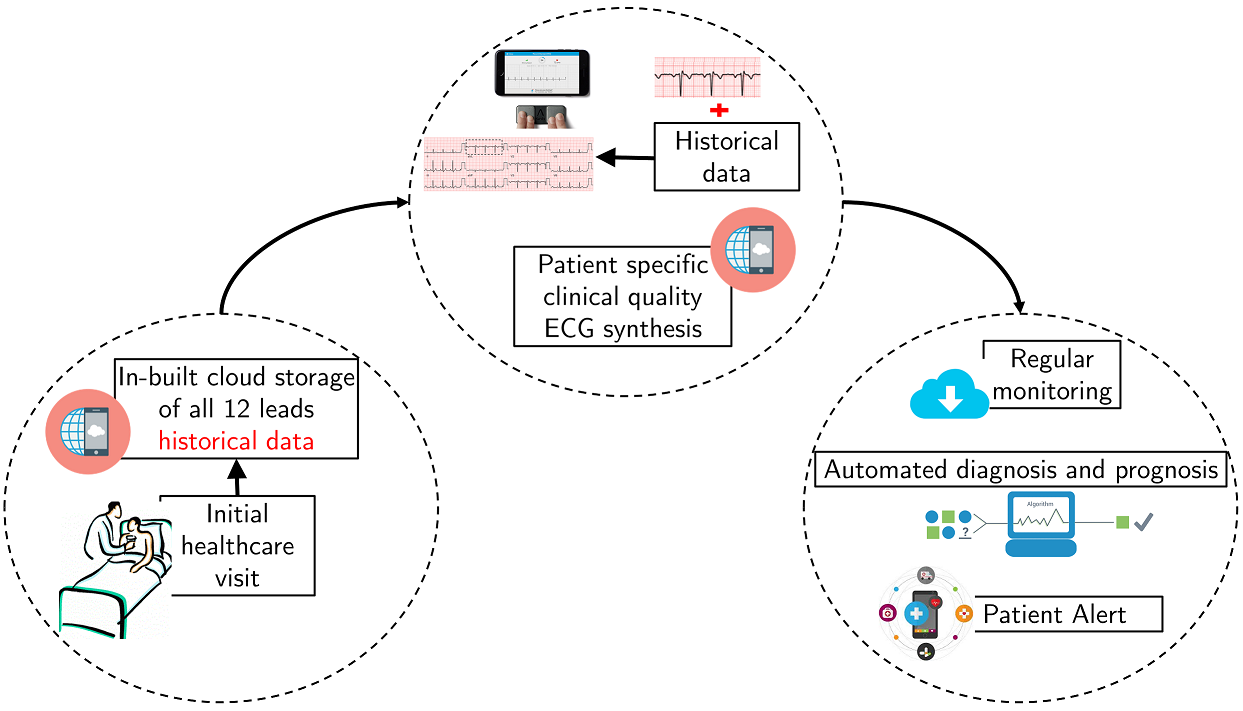}
		\caption{Envisioned eventual goal of developing an integrated POC system using a single-lead handheld device. This system can be employed for acute CVD diagnosis and management using the synthesized 12-lead ECG.}
	\label{envision}
\end{figure}

Synthesis of the missing leads is to make the standard ECG available to the cardiologist and primary physicians as they are trained to parse 12-lead signal. Although parsing a single-lead  is much easier, it does not give sufficient information for detecting situations leading to sudden cardiac death. The POC handheld ECG devices have significantly improved the POC diagnosis situations for the acute cardiac patients especially for those with limited mobility and comorbidity.

We envision the synthesis of 12-lead from handheld ECG device to be part of a bigger POC system as shown in Figure~\ref{envision}. At the time of handheld ECG system prescription to a subject, their simultaneous 12-lead ECG recording can be uploaded to a secure cloud-based server. Every time the subject uses the handheld device and records a single-lead ECG, that ``current" recording can be used to synthesize all the missing 11 lead signals at that time. This generates the 12-lead standard ECG and updates the historical data. Consequently, this 12-lead ECG can be used for automated diagnosis and can be integrated into a user app to provide alert for adverse events. A series of historical data being stored on the cloud server can also be accessed by the healthcare provider for monitoring the long-term health of the subject and provide a more robust diagnosis.

The proposed methodology is simple enough to be implemented in a real-time system. Nonetheless, the main thrust of our methodology relies on accurate detection of fiducial beats. Although our fiducial marker detection scheme was nearly accurate, it can result in delays in the real-time diagnosis system. Hence, one of the possible future work directions could be incorporating a faster beat detection algorithm and to make the overall algorithm faster. Further, an improved lag prediction algorithm can significantly improve the sensitivity of the synthesis. Ultimately, the aim is to use these at-home synthesized 12-lead ECG for near clinical equivalent ECG screening and diagnosing a broader set of cardiac abnormalities, not limited to arrhythmia detection.

\section*{Acknowledgment}
{\color{black} We would like to thank the U.S. National Science Foundation under Grant NSF PFI: AIR-TT 1543226 and NSF CMMI 1301439 for their support in this work. 
We also want to sincerely thank the Heart, Artery, and Veins Center of Fresno, California for providing the Kardia-AliveCor\textregistered~and the paper ECG data. }

\bibliographystyle{IEEEtran}
\bibliography{ECG_arXiv}

\end{document}